\documentclass[preprint,numbers,sort&compress,12pt]{elsarticle}

\usepackage{natbib}

\usepackage{amsfonts}
\usepackage{amssymb}
\usepackage{amsmath}

\usepackage{amssymb}
\usepackage{amsthm}

\newcommand{\bsym}[1]{\ensuremath{\boldsymbol{#1}}}
\newcommand{\w}{\ensuremath{\boldsymbol{\wedge}}}

\journal{ArXiv}

\begin{document}

\begin{frontmatter}



\title{Reduction of $T^*SE(3)$ to the Poisson structure for a symmetric top}


\author[sszub]{Stanislav S. Zub}
\ead{stah@univ.kiev.ua}
\author[sizub]{Sergiy I. Zub}

\address[sszub]{Faculty of Cybernetics. Taras Shevchenko National University of Kyiv,
Glushkov boul., 2, corps 6., Kyiv, Ukraine 03680}
\address[sizub]{Institute of Metrology,
Mironositskaya st., 42, Kharkiv, Ukraine 61002}



\end{frontmatter}


\section{The Hamiltonian formalism on $T^*SO(3)$}
\label{hformso3}

{\bf Representation of the right trivialization of $T^*SO(3)$}

Here we show some useful relations for the group $SO(3)$ and its cotangent bundle,
many of them can be found in \cite{AbrMar02,MarRat98}.

Group $SO(3)$ is formed by the orthogonal, unimodular matrices $\mathbf{R}$ i.e.
$\mathbf{R}^T = \mathbf{R}^{-1}$, $\rm det(\mathbf{R})= 1$.
Accordingly the Lie algebra $\bsym{so(3)}$ is formed by $3\times 3$-antisymmetric matrix
with Lie bracket in the form of matrix commutator.

Let us consider a vector space isomorphism $\hat{}:\mathbb{R}^3\rightarrow\bsym{so(3)}$
such that \cite[p. 285]{MarRat98}:
$\widehat{\bsym{\xi}}_{kl} = -\varepsilon_{ikl}\xi_i$, $\xi_i = -\frac12 \varepsilon_{ikl} \widehat{\bsym{\xi}}_{kl}$,
where $\varepsilon_{ikl}$ --- Levi-Civita symbol. Then the next equations will be satisfied
\begin{equation}
\begin{cases}
   \widehat{\bsym{\xi}}\bsym{\eta} = \bsym{\xi}\times \bsym{\eta};\\
   [\widehat{\bsym{\xi}},\widehat{\bsym{\eta}}] =
   \widehat{\bsym{\xi}}\widehat{\bsym{\eta}} - \widehat{\bsym{\eta}}\widehat{\bsym{\xi}} =
   \widehat{\bsym{\xi}\times \bsym{\eta}};\\
   \langle \bsym{\xi},\bsym{\eta} \rangle
   = -\frac12 {\rm tr}(\widehat{\bsym{\xi}}\widehat{\bsym{\eta}});\\
   \mathbf{B}\widehat{\bsym{\xi}}\mathbf{B}^{-1} = \widehat{\mathbf{B}\bsym{\xi}}
\end{cases}
\label{A1}
\end{equation}

The scalar product introduced above allows us to state the equivalence
of the Lie algebra and its dual space $\bsym{so(3)}^*\simeq\bsym{so(3)}$.
Symbol ``$\simeq$'' defines diffeomorphism.
Usually, the diffeomorphisms used below have the simple group-theoretical
or differential-geometric sense that is explained in literature. 

In the representation of {\it right trivialization} which is shown \cite{LeRaSiMa92}
and \cite[p. 314]{AbrMar02} to correspond to the {\it inertial frame of reference},
we have:\\
$T SO(3)\simeq SO(3)\times\bsym{so(3)}$, $T^*SO(3)\simeq SO(3)\times\bsym{so(3)}^*$.

Then right and left actions of the group $SO(3)$ on $T^*SO(3)$ have the form
\begin{equation}
\begin{cases}
   R_{\mathbf{B}}: (\mathbf{R},\widehat{\bsym{\pi}})\in T^*SO(3)
      \rightarrow(\mathbf{R}\mathbf{B},\widehat{\bsym{\pi}}), \quad \mathbf{B}\in SO(3);\\
   L_{\mathbf{B}^{-1}}: (\mathbf{R},\widehat{\bsym{\pi}})\in T^*SO(3)
      \rightarrow(\mathbf{B}^{-1}\mathbf{R},\mathbf{B}\widehat{\bsym{\pi}}\mathbf{B}^{-1})
   = (\mathbf{B}^{-1}\mathbf{R},{\rm Ad}^*_{\mathbf{B}^{-1}}\widehat{\bsym{\pi}})
\end{cases}
\label{A2}
\end{equation}

{\bf Symplectic and Poisson structures on $T^*SO(3)$} \cite{AbrMar02,LeRaSiMa92}

Liouville form on $T^*SO(3)\simeq SO(3)\times\bsym{so(3)}^*$ has the form
\begin{equation}
   \Theta^{T^*SO(3)}_{|(\mathbf{R},\bsym{\pi})}
   = -\frac12 {\rm tr}(\widehat{\bsym{\pi}}\widehat{\bsym{\delta R}})
   = \pi_i\delta R^i,
\label{A3}
\end{equation}
where $\widehat{\bsym{\delta R}} = (d\mathbf{R})\mathbf{R}^{-1}$ --- right-invariant Maurer-Cartan 1-form.

Then by using the Maurer-Cartan equation \cite[p. 276]{MarRat98}, for the canonical symplectic 2-form we have
\begin{equation}
   \Omega^{T^*SO(3)}_{can}  = -d\Theta^{T^*SO(3)} = {\delta R}^i\w d\pi_i - \pi_i [\delta R,\delta R]^i =
\label{A4}
\end{equation}
\[ = -\frac12\varepsilon_{ijk}\delta R_{jk}\w d \pi_i
   + \frac12 \pi_i\varepsilon_{ijk}\delta R_{js}\w \delta R_{sk}
\]

Any given symplectic structure $\Omega$ defines a Poisson structure on the same manifold as follows
\begin{equation}
   \{F, G\}(z) = \Omega(\xi_F(z),\xi_G(z))
   = \partial_{\xi_G}F = -\partial_{\xi_F}G,
\label{A5}
\end{equation}
where equation $i_{\xi_G}\Omega = dG$ will be satisfied for the vector field $\xi_G$.

By considering the elements of the matrix $\mathbf{R}$ and components of the momentum $\bsym{\pi}$
as the dynamic variables on $T^*SO(3)$, we can obtain the following set of Poisson brackets,
that completely defines a Poisson structure on $T^*SO(3)$:
\begin{equation}
   \{R_{ij}, R_{kl}\} = 0, \quad
   \{\pi_i, R_{jk}\} = \varepsilon_{ijl}R_{lk}, \quad
   \{\pi_i, \pi_j\} = \varepsilon_{ijl}\pi_l.
\label{A6}
\end{equation}

Note that in the inertial system the Poisson brackets for the matrix elements $\mathbf {R}$
are grouped in columns, for example, Poisson brackets for the elements of the 3d-column is expressed
in terms of the elements of the 3d-column only.

\section{Reduction of the Poisson structure for a symmetric top}
\label{redforsymtop}

Poisson structure (\ref{A6}) is invariant under right translations of constant matrix $\mathbf{B} \in SO(3)$:
\begin{equation}
   \{(R B)_{ij}, (R B)_{kl}\} = 0, \quad
   \{\pi_i, (R B)_{jn}\} = \varepsilon_{ijl}(R B)_{ln}, \quad
   \{\pi_i, \pi_j\} = \varepsilon_{ijl}\pi_l.
\label{B1}
\end{equation}

The subgroup $S^1~\in~SO(3)$, where $S^1 = \{ \mathbf{Z}\in SO(3): Z_{i3}= \delta_{i3} \}$
is of interest in connection with the symmetric top (it is assumed that the axis of the body symmetry
is directed along the vector $\bsym{E}_3$).
Then $(\mathbf{R}\mathbf{Z})_{i3} = R_{ij}Z_{j3} = R_{i3}$,
i.e. 3d-row of the matrix $\mathbf{R}$ remains invariant under right translations
that corresponds to the subgroup $S^1$.

Let's consider the projection $SO(3)$ on the sphere $S^2$ (as a set of unit vectors $\bsym{\nu}^2=1$)
\begin{equation}
   \tau: \mathbf{R}\mapsto \bsym{\nu} = R_{i3}\bsym{e}_i, \quad \nu_i = R_{i3}.
\label{B2}
\end{equation}

This projection generates a map
\begin{equation}
   \tilde{\tau}: T^*SO(3)\ni (\mathbf{R},\bsym{\pi})
   \mapsto (\tau(\mathbf{R}),\bsym{\pi}) = (\bsym{\nu},\bsym{\pi})\in W_1,
\label{B3}
\end{equation}
where $W_1\simeq S^2\times \bsym{so(3)}^*\subset \mathbb{R}^3\times \bsym{so(3)}^*\simeq\bsym{se(3)}^* = \mathbb{R}^3\circledS\bsym{so(3)}^*$.

On $\bsym{se(3)}^*$, as on any space that is dual to the Lie algebra
there exists a canonical structure of  Lie-Poisson \cite[p. 425]{MarRat98}.

In this case \cite[pp. 491,367]{MarRat98} this structure is determined by the following Poisson brackets
\begin{equation}
   \{\nu_i, \nu_k\} = 0, \quad
   \{\pi_i, \nu_j\} = \varepsilon_{ijl}\nu_l, \quad
   \{\pi_i, \pi_j\} = \varepsilon_{ijl}\pi_l.
\label{B4}
\end{equation}

Comparing (\ref{B4}) with (\ref{A6}) shows that a surjective mapping $\tilde{\tau}$ is poissonian.

Thus the conditions of {\it Theorem 10.5.1} \cite[p. 355]{MarRat98} are satisfied.
And hence,
\begin{equation}
   W_1 \simeq T^*SO(3)/ S^1
\label{B5}
\end{equation}

If Hamiltonian $H$ on $T^*SO(3)$ $S^1$ is invariant then on $W_1$
there is a Hamiltonian $h$ such that $H=h\circ \tilde{\tau}$
and the trajectories of the dynamic system with Hamiltonian $H$
$\tilde{\tau}$-associated with the trajectories for the Hamiltonian $h$ \cite[p. 355]{MarRat98}.

\section{The structure of symplectic leaves $\bsym{so(3)}^*$}
\label{symleavesso3}

{\bf The levels of Casimir functions as the orbits of the coadjoint representation}

As it was proven above
\begin{equation}
   W_1 \simeq T^*SO(3)/ S^1 = \{(\bsym{\nu},\bsym{\pi})\in \bsym{se(3)}^*: \bsym{\nu}^2 = 1\}.
\label{C1}
\end{equation}

Coadjoint action of the group $SE(3)$ on $\bsym{se(3)}^*$ has the form (see (14.7.10) \cite{MarRat98})
\begin{equation}
   Ad^*_{(\bsym{a},\mathbf{A})^{-1})}(\bsym{\nu},\bsym{\pi})
   = (\mathbf{A}[\bsym{\nu}],\bsym{a}\times \mathbf{A}[\bsym{\nu}] + \mathbf{A}[\bsym{\pi}]).
\label{C2}
\end{equation}

Let's consider the functions
\begin{equation}
   C_1(\bsym{\nu},\bsym{\pi}) = \bsym{\nu}^2, \quad
   C_2(\bsym{\nu},\bsym{\pi}) = \bsym{\nu}\cdot\bsym{\pi}.
\label{C3}
\end{equation}

{\bf Proposition 1}. {\it Functions (\ref{C3}) are invariant relative to the coadjoint action (\ref{C2}),
so on the basis of the proposition 12.6.1 \cite[p. 421]{MarRat98} we can argue that $C_1,C_2$
are the Casimir functions on $\bsym{se(3)}^*$}.

Thus, $W_1$ is submanifold in $\bsym{se(3)}^*$ that is determined by the level of the Casimir function $C_1$
on $\bsym{se(3)}^*$, and hence, it is {\it Poisson submanifold} $\bsym{se(3)}^*$.
Then the symplectic leaves, on which $W_1$ stratifies is also the symplectic leaves of the Lie-Poisson structure on $\bsym{se(3)}^*$.

Let's consider the symplectic leaves of the Lie-Poisson structure on $\bsym{se(3)}^*$.
From {\it Corollary 14.4.3} \cite[p. 477]{MarRat98} follows
that they are the orbits of the coadjoint action of $SE(3)$.
The structure of these orbits was studied in \cite[\S14.7]{MarRat98}
and formulated in {\it Theorem 4.4.1} (see \cite[p. 142]{MarMisOrtPerRat07}).

The trajectory (accumulation curve) of any Hamiltonian field
emanating from a point $(\bsym{\nu}_0,\bsym{\pi}_0)$
remains on a joint level of the Casimir functions:
\begin{equation}
   \textsl{L}_{(\bsym{\nu}_0,\bsym{\pi}_0)}
   = \{(\bsym{\nu},\bsym{\pi}):C_1(\bsym{\nu},\bsym{\pi}) = \bsym{\nu}_0^2,
   C_2(\bsym{\nu},\bsym{\pi}) = \bsym{\nu}_0\cdot\bsym{\pi}_0\}\subset\bsym{se3}^*
\label{C4}
\end{equation}

{\bf Proposition 2}. {\it A joint level $\textsl{L}_{(\bsym{\nu}_0,\bsym{\pi}_0)}$
of the Casimir functions~(\ref{C3}) with $\bsym{\nu}_0\neq 0$ is the orbit of
$\textsl{O}_{(\bsym{\nu}_0,\bsym{\pi}_0)}$ of the coadjoint representation of the group $SE(3)$}.

Indeed, by following formula (\ref{C2}) it is easy to give a direct proof of transitivity
of the action $SE(3)$ on $\textsl{L}_{(\bsym{\nu}_0,\bsym{\pi}_0)}$.

{\bf 2-form of Kirillov-Kostant-Souriau on the symplectic leaves $\bsym{so(3)}^*$}

The symplectic structure $\Omega^G_{KKS}$ named after Kirillov-Kostant-Souriau
is defined on the orbits of the coadjoint representation of the Lie group $G$.

This structure may be obtained by reduction from the canonical structure
on $(T^*G,\Omega^{T^*G}_{can})$ and agrees with the Lie-Poisson structure on $\bsym{g}^*$
as follows from {\it Theorem 14.4.1} in \cite[p. 475]{MarRat98}.

Structure of the orbits $\textsl{O}_{(\bsym{\nu}_0,\bsym{\pi}_0)}$ on $\bsym{se(3)}^*$,
where $\bsym{\nu}_0\neq 0$ is ascertained by propositions 2,3 with section 14.7 \cite{MarRat98}
and with example 4.4 of the book \cite{MarMisOrtPerRat07}.

{\it Remark}. {\it Type1} orbits from section 14.7 \cite{MarRat98} are irrelevant to the dynamics
of a symmetric top because $\bsym{\nu}^2 = 1$.
In addition, results of the {\it Type2} are completely covered by results {\it Type3}.
Thus it suffices to use {\it Theorem 4.4.1} of the book \cite{MarMisOrtPerRat07} for {\it Type3}.

{\bf Proposition 3}. {\it Orbit $\textsl{O}_{(\bsym{\nu}_0,\bsym{\pi}_0)}$ with $\bsym{\nu}^2_0=1$
is diffeomorphic to the cotangent bundle of the sphere, and the symplectic form on the orbit differs
from the canonical symplectic form on the cotangent bundle of the sphere in so-called magnetic term.
\begin{equation}
\begin{cases}
   \textsl{O}_{(\bsym{\nu}_0,\bsym{\pi}_0)} \simeq T^*S^2, \quad \bsym{\nu}^2_0=1;\\
   \Omega^{\textsl{O}_{(\bsym{\nu}_0,\bsym{\pi}_0)}}_{KKS}
   = \Omega^{T^*S^2}_{can}-\rho^*\textsl{B};\\
   \textsl{B}_{|\bsym{\nu}}(\bsym{\xi}\times \bsym{\nu}, \bsym{\eta}\times \bsym{\nu})
   = -C_2(\bsym{\nu}_0,\bsym{\pi}_0)\langle \bsym{\xi}\times \bsym{\eta},\bsym{\nu}\rangle
\end{cases}
\label{C5}
\end{equation}
where $\rho:T^*S^2\rightarrow S^2$ is the cotangent bundle projection,
$\textsl{B}$ --- 2-form on the sphere,
$\bsym{\nu}\in S^2$, $\bsym{\xi},\bsym{\eta}\in\mathbb{R}^3$}.

As we can see from the 3-rd line (\ref{C5}) the magnetic term is $\textsl{B}\rightarrow0$
with  $\bsym{\pi}_0\rightarrow0$.
Therefore, the 2-nd type of the orbits in {\it Theorem 4.4.1}
(see \cite[p. 142]{MarMisOrtPerRat07}) is a special case of the 3-rd type.
Note also that 2-nd type of the orbits leads to the unstable orbital motion in the Orbitron task.

\section{Reduction of $T^*SE(3)$ to the Poisson structure for a symmetric top}
\label{redse3tosymtop}

{\bf Group $SE(3)$ as a common configuration space for rigid body dynamics}

The common description of the Hamiltonian mechanics of the rigid body is $(T^*SE(3),\Omega^{T^*SE(3)}_{can},H)$,
where $\Omega^{T^*SE(3)}_{can} = -d \Theta^{T^*SE(3)}$,
and $\Theta^{T^*SE(3)}$ --- Liouville 1-form on $T^*SE(3)$.

Let's consider the body frame be equal to the orthonormal frame (triad of $\{\vec{E}_i\}$)
with a zero point in the center of mass of the rigid body
and the unit vectors directed along the principal axes of inertia tensor of the body
with the same orientation as in the spatial frame $\{\vec{e}_i\}$.

Let's introduce the transition matrix $\mathbf{R}$ from the spatial frame ($\{\vec{e}_i\}$)
to the $\{\vec{E}_i\}$ body frame.
\begin{equation}
   \vec{E}_k = R_{ik}\vec{e}_i, \quad
   R_{ji} = \langle \vec{E}_i, \vec{e}_j\rangle,
\label{D1}
\end{equation}
where matrix $\mathbf{R}$ has the following properties:
$\mathbf{R}^T = \mathbf{R}^{-1}, \rm det(\mathbf{R})= 1$.

Elements $T^*SE(3)$ in the inertial frame of reference is defined by four values
$((\bsym{x},\mathbf{R}),(\bsym{p},\bsym{\pi}))$,
where $\bsym{p}$ --- momentum of translation motion
and \\$\bsym{\pi}$ --- intrinsic angular momentum of the rigid body.

{\bf The canonical Poisson structure on $T^*SE(3)$}

In the inertial frame of reference (right trivialization) a canonical symplectic structure on a manifold
\begin{equation}
   T^*SE(3) \simeq T^*(\mathbb{R}^3)\times T^*(SO(3))
   \simeq T^*(\mathbb{R}^3)\times SO(3)\times \bsym{so(3)^*}
\label{D2}
\end{equation}
is a direct product \cite[p. 81--82]{Souriau97},
where the translational and rotational degrees of freedom are separated
from each other in the symplectic and Poisson structures
\begin{equation}
   \Omega^{T^*SE(3)}_{can} = \Omega^{T^*R^3}_{can} + \Omega^{T^*SO(3)}_{can}
\label{D3}
\end{equation}
\[ = d x^i\w d p_i -\frac12\varepsilon_{ijk}\delta A_{jk}\w d \pi_i
   + \frac12 \pi_i\varepsilon_{ijk}\delta A_{js}\w \delta A_{sk}
\]

In the {\it inertial frame of reference} non-zero Poisson brackets
that correspond to $\Omega^{T^*SE(3)}_{can}$, have the form
\begin{equation}
   \{x_i, p_j\} = \delta_{ij}, \quad
   \{\pi_i, \pi_j\} = \varepsilon_{ijl}\pi_l, \quad
   \{\pi_i, R_{jk}\} = \varepsilon_{ijl}R_{lk}.
\label{D4}
\end{equation}

{\bf Reduction $T^*SE(3)$ to $T^*SE(3)/ S^1$}

As in the case of $T^*SO(3)$, the Poisson brackets for $T^*SE(3)$ are invariant for right translations
\begin{equation}
   R_{\mathbf{B}}:((\bsym{x},\mathbf{R}),(\bsym{p},\bsym{\pi}))
   \mapsto ((\bsym{x},\mathbf{R}\mathbf{B}),(\bsym{p},\bsym{\pi})),
   \quad \mathbf{B} \in SO(3)
\label{D5}
\end{equation}

As we can seen from (\ref{D5}), variables $\bsym{x},\bsym{p}$ for translational degrees of freedom
are not subjected to transformation, and play a passive role
in the  procedure of reduction and they can be omited in further discussion.

In general case the kinetic energy of the rigid body is left-invariant,
but not right-invariant. However, for the symmetric top, the system is right translation invariant to $S^1 \in SO(3)$,
where the group $S^1$ the rigid body symmetry (assuming,
that the axis of the body symmetry is directed along the vector $\bsym{E}_3$).

As in {\it Section 2} the conditions of {\it Theorem 10.5.1} \cite[p. 355]{MarRat98} are satisfied,
and therefore, for the Poisson manifold $P_1$ we obtain
\begin{equation}
   P_1 = T^*SE(3)/ S^1 \simeq T^*\mathbb{R}^3\times W_1,
\label{D6}
\end{equation}
with such Poisson brackets
\begin{equation}
   \{x_i, p_j\} = \delta_{ij}, \quad
   \{\nu_i, \nu_k\} = 0, \quad
   \{\pi_i, \nu_j\} = \varepsilon_{ijl}\nu_l, \quad
   \{\pi_i, \pi_j\} = \varepsilon_{ijl}\pi_l.
\label{D7}
\end{equation}

In order to carry out the reduction of the system in full,
the standard Hamiltonian of a rigid body, namely,
the contribution of the kinetic energy of proper rotation
must be converted into the inertial frame of reference.
This is not difficult to perform for the symmetric top,
where two momenta of inertia are equal in the body frame.

By using $I_1 = I_2 = I_\perp$, after some transformations we obtain
\begin{equation}
   T_{spin}(((\bsym{x}, \mathbf{R}), (\bsym{p},\bsym{\pi})) )
   = \frac1{2 I_1} \bsym{\pi}^2
   + \left(\frac1{2 I_3} - \frac1{2 I_1}\right)\langle\bsym{\nu}, \bsym{\pi}\rangle^2
   + V(\bsym{x},\bsym{\nu}).
\label{D8}
\end{equation}

Thus, after discarding Casimir function
$\left(\frac1{2 I_3} - \frac1{2 I_1}\right)\langle\bsym{\nu}, \bsym{\pi}\rangle^2$
the Hamiltonian of the symmetric top takes the form
\begin{equation}
   h(((\bsym{x}, \bsym{\nu}), (\bsym{p},\bsym{\pi})) )
   = \frac1{2 M} \bsym{p}^2 + \frac1{2 I_1} \bsym{\pi}^2
   + V(\bsym{x},\bsym{\nu}),
\label{D9}
\end{equation}
where $M$ --- mass of the rigid body

Hamiltonian $h$ depends only on the dynamic variables on $P_1 \subset P$.

The dynamic system is finally reduced to the $(P_1,\{\cdot,\cdot\},h)$,
because from {\it Theorem 10.5.1} \cite[p. 355]{MarRat98}
the dynamic trajectories of the reduced system are the projections
of the dynamic trajectories of the original system via Poisson mapping.
\begin{equation}
   T^*SE(3)\ni ((\bsym{x},\mathbf{R}),(\bsym{p},\bsym{\pi}))
   \mapsto ((\bsym{x},\tau(\mathbf{R})),(\bsym{p},\bsym{\pi}))
   \in P_1 \subset P
\label{D10}
\end{equation}

With regard to the structure of the symplectic leaves
$\textsl{L}^{P_1}_{(\bsym{\nu}_0,\bsym{\pi}_0)}$ of the Poisson manifold $P_1$,
taking into consideration the results in {\it Section 3}, we have
\begin{equation}
   \textsl{L}^{P_1}_{(\bsym{\nu}_0,\bsym{\pi}_0)}
   = T^*\mathbb{R}^3\times \textsl{O}_{(\bsym{\nu}_0,\bsym{\pi}_0)}
\label{D11}
\end{equation}
see {\bf Proposition 3}.



\bibliographystyle{elsarticle-num}
\bibliography{my-bibdb}







\end{document}